# DYNAMICAL CHARACTERIZATION OF THE LAST PROLONGED SOLAR MINIMA

Rodolfo Gustavo Cionco[a,*] and Rosa Hilda Compagnucci[b]

[a] *Universidad Tecnológica Nacional (UTN), Facultad Regional San Nicolás, Colón 332, San Nicolás de los Arroyos, (2900), Argentina, gcionco@frsn.utn.edu.ar*

[b] *Universidad de Buenos Aires, Departamento de Ciencias de la Atmósfera y los* Océanos, FCEN/UBA, and *CONICET, Ciudad Universitaria, Pabellón II; CABA, (1428),* Argentina, rhc@fcen.uba.ar

\* Corresponding author: Dr. Rodolfo Gustavo Cionco, gcionco@frsn.utn.edu.ar , Colón 332, San Nicolás de los Arroyos, (2900), Buenos Aires, Argentina. Tel.-FAX #54 - 0336- 4420830 /4425266.

Abstract

The planetary hypothesis of the solar cycle is an old idea in which the gravitational influence of the planets has a non-negligible effect on the causes of the solar magnetic cycle. The advance of this hypothesis is based on phenomenological correlations between dynamical parameters of the Sun´s movement around the barycentre of the Solar System and sunspots time series; and more especially, identifying relationships linking solar barycentric dynamics with prolonged minima (especially Grand Minima events). However, at present there is no clear physical mechanism relating these phenomena. The possible celestial influence on solar cycle modulation is of great importance not only in solar physics but also in Earth sciences, because prolonged solar minima have associated important climatic and telluric variations, in particular, during the Maunder and Dalton Minima. In this work we looked for a possible causal link in relation with solar barycentric dynamics and prolonged minima events. We searched for particular changes in the Sun's acceleration and concentrated on long-term variations of the solar cycle. We show how the orbital angular momentum of the Sun evolves and how the inclination of the solar barycentric orbit varies during the epochs of orbital retrogressions. In particular, at these moments, the radial component of the Sun's acceleration (i.e., in the barycentre-Sun direction) had an exceptional magnitude. These radial impulses occurred at the very beginning of the Maunder Minimum, during the Dalton Minimum and also at the maximum of cycle 22 before the present extended minimum. We also found a strong correlation between the planetary torque and the observed sunspots international number around that maximum. We apply our results in a novel theory of Sun-planets interaction that it is sensitive to Sun barycentric dynamics and found a very important effect on the Sun´s capability of storing hypothetical reservoirs of potential energy that could be released by internal flows and might be related to the solar cycle. This process begins about 40 years *before* the solar angular momentum inversions, i.e., before Maunder Minimum, Dalton Minimum, and before the present extended minimum. Our conclusions suggest a dynamical characterization of peculiar prolonged solar minima. We discuss the possible implications of these results for the solar cycle including the present extended minimum.

## 1. Introduction

The *planetary hypothesis* of the solar cycle is an old idea in which the planets of the solar system have some influence on solar activity, mainly seen in modulations in the number of sunspots. Studies disseminated elsewhere have treated this issue at different levels. Several of these works were published in important astronomy-astrophysics-geophysics journals and concentrated on studying this possible Sun-planets relationship (e. g., Wolf, 1859; Brown, 1900; Jose, 1965; Wood and Wood, 1965; Lansdscheidt, 1999; Fairbridge and Shirley, 1987; Charvátová, 1997, 2000, 2009; Grandpierre, 1996, Juckett 2000, 2003; Javaraiah, 2005; Shirley, 2006; Wolf and Patrone, 2010; Perryman and Schulze-Hartung, 2011). Most of these contributions relied on the study of the planar projection of the movement of the Sun, as a point mass body, around the centre of mass (*the barycentre*) of the Solar System, the so called *solar inertial motion* (SIM), and to assess correlations between features of SIM and solar activity.

Over the years, several lines of inquiry were proposed to explain these relationships by which a putative mechanism (i.e., a working mechanism linking solar orbital dynamics with internal dynamics of the Sun), can produce sporadic or cyclical modifications on solar magnetism. These ideas included the raising of the tides by planets (see e.g., Takahashi, 1968; Wood, 1972) which has been largely discredited (e.g., de Jager and Versteegh, 2005); velocity and acceleration variations of the solar motion, (Wood and Wood, 1965; Fairbridge and Shirley, 1987, Charvátová , 2009); and a third line of inquiry involves the orbital angular momentum and its derivatives, which has received the most attention (Jose, 1965; Jucket, 2000; Javaraiah, 2005; Shirley, 2006; Perryman and Schulze-Hartung, 2011).

The movement of the Sun around the barycentre is a consequence of the *N*-body dynamics operating in the Solar System. This gravitational motion of the Sun's centre of mass produces a complex orbit which, due to the massive planet Jupiter (which has a mass ≈ $10^{-3}$ Ms, Ms = 1 solar mass, approximately 70% of the whole planetary mass), should be confined to a zone in the order of ≈ $10^{-3}$ $a_J$ ≈ 5.2 x $10^{-3}$ AU ($a_J$ is Jupiter´s orbit semi-major axis) around the barycentre of the solar system, as may be easily checked by the reader by consulting the works of, for example, Jose (1965) or Charvátová (2009).

In his pioneering work, Jose (1965) used the giant planets orbital integrations of 1951 to compute the SIM between 1653 and 2060 A. D. He pointed out an apparent synchronization between dynamical parameters of SIM (modulus of the orbital angular momentum, *L,* and its derivative, d*L*/d*t* the so-called planetary "torque") and sunspot series, but using suppositions not based on any physics (i.e., by changing cycle polarity at strategic times). He used three dimensional (3D) positions of the planets, and reported sporadic angular momentum inversions, that is, *retrograde* solar motion around the barycentre in 1811 and 1990; but he did not mention any noticeable effect on solar osculating orbit (OSO) at these times. He found that dynamical variables such as *r* (Sun barycentric distance), *L* and d*L*/d*t* repeat approximately after ~179 yr. This period has a close connection with the planetary positions of the giant planets (the synodic period of Uranus and Neptune is ~ 171 yr), i. e., with giant planets quasi-alignments (quasi- conjunctions of Jupiter with the other three giants). Authors such as Fairbridge and Shirley (1987), Charvátová (1997, 2000) and Landscheidt (1988, 1999), have focused on the relationship between SIM and extreme values in sunspots series, especially related to the prolonged solar minima of the last millennium, known as Grand Minima (GM) events.
The GM events are prolonged and pronounced minima of different duration in the Sun´s magnetic activity, seen as deep drops in observed and inferred sunspot numbers (e.g., Eddy, 1976; Stuiver and Quay, 1981; Solanki et al., 2004; Usoskin, 2008). The most studied of these events, the Maunder Minimum (MM, 1645-1712) and Dalton Minimum (DM, 1790- 1830) were periods of great decrease in sunspots number; in addition, these epochs have been related with a noticeable mean temperature drop; as has been confirmed by historical and archaeological data (Lamb, 1995) and recently by proxy-data reconstruction of the mean temperature anomalies over global and north hemispheric regions (Lohele and McCulloch, 2008; Mann et al., 2009). The occurrences of GM events have been especially linked with the planetary hypothesis through SIM. Fairbridge and Shirley (1987), were able to show that only at times of the main GM events, a substitute of the apsidal axis of the solar barycentric orbit has had strong oscillations, but in their work there is no hypothesis related to any particular forcing mechanism in connection with this phenomenon. They also found a strong oscillation in 1990, so the authors, extrapolating these findings, argued for a new imminent GM after that epoch. In other works, Landscheidt (1988, 1999) quantified changes in planetary torque and relates d*L*/d*t* impulses at epochs of closest approaches Sun-barycentre with GM. Landscheidt argued that solar plasma trails can be released in these impulsive events but his presentation lacks any physical mechanism that explains this. He also found a very profound change in *L* around 1990 and this was used by the author to argue that an imminent minimum in solar activity would result.

Charvátová (2000, 2009) showed that the previously mentioned quasi-apsidal oscillations are seen in SIM as non-regular or "disordered-like" solar orbits; but in the lapses between the GM events the solar orbit is symmetric "tri-foliar-like". She notes that the Sun enters in tri-foliar orbits with a periodicity of ~179 yr and called these two types of solar orbits "chaotic" and "regular". As a comment, it is important to note that macroscopic chaos in planetary-sized bodies can be seen after several million years (Laskar, 1994). Charvátová through these qualitative similarities in SIM and sunspot series related to GM, especially DM, hypothesized that an imminent new GM event might occur after the cycle 22.

In general, all these authors related directly or indirectly prolonged minima events with planetary alignment (note that Fairbridge and Shirkey, 1987, had not taken into account the DM as a GM event) and many of them refer to a possible new GM before the cycle 22.

All of these conclusions are very interesting, especially because the current unexpected long minimum of solar activity has attracted attention with respect to the possibility of developing a new GM event in the next decades (e.g., Badalyan et al., 2001; Livingston and Penn, 2009; Russell et al., 2009; Tan, 2010; Ahluwalia and Ygbuhay, 2012). Although the "precise" classification of a prolonged minimum as a GM event is very important, this minimum has conspicuous properties for which it has been called "exceptional" (e.g., Gibson et al., 2009) and it is important by itself. These properties must also be discussed with other general properties of GM events, such as their possible cyclicity or stochastic nature, which are under debate.

Although giant planets quasi-conjunctions are events that generate very particular gravitational forces, so far, no one has shown what this particular forcing is like. All the aforementioned estimates are based on crude parameterizations of Sun barycentric dynamics; nevertheless, they are quite remarkable in light of the lack of a distinctive planetary forcing in Sun barycentric dynamics linking planetary movements and solar magnetic activity at GM events epochs. GM events, have not only been associated with systematic drops in sunspots number, but also with sporadic anomalies in the numbers; for example, Gnevyshev-Ohl (G-O) odd-even rule violations (G-O rule states that the sum of sunspot numbers over an odd-numbered solar cycle exceeds that of its preceding even -numbered cycle), the last one occurred in 1990, at the maximum of cycle 22 (Javaraiah, 2005). Therefore, these lines of evidence strongly suggest that a possible relationship between planetary movements and long term solar cycle is operating. Even more interesting, a relationship between SIM and terrestrial climate was reported. Indeed, notwithstanding the complexities and relationships between solar irradiance variations, solar modulated cosmic-rays flux and climate (Gray et al., 2010, and references therein), a possible SIM-Earth climate connection has been shown (e. g., Landschidt, 1988; Charvátová, 1997; Scafetta, 2010) and a very specific *planetary signal* in Earth climate has been reported (Charvátová and Štreštík, 2004; Scafetta, 2010). A varying Sun, in a broad sense (taking into account internal and external causes) would influence the climate by means of several complicated mechanisms and feedbacks.

Several works have showed phenomenological relationships between solar cycle and seismic and volcanic activity (e. g., Odintsov et al., 2007; Khain and Khalilov, 2008). Moreover, volcanic activity seems to be enhanced at epochs of GM events especially after 1632 and 1811, as has been shown by Lamb (1983); Khain and Kalilov (2008); Wagner and Zorita (2005). All this body of evidence strongly suggests a teleconnection between Earth climate-telluric activity and solar cycle, especially at GM events.

The mechanism called upon to explain the solar activity, of which sunspots are the most evident manifestation, is the solar dynamo (e. g., Charbonneau, 2010; Karak, 2010, and references therein). The solar dynamo theory can explain at least the most basic features of the solar cycle. Although the predictions of dynamo theory do not agree with the phenomenology of the solar cycle very well, the Schawbe (11 yr) period certainly can be simulated using mean-field dynamo with adequate external parameterizations, and GM events like MM can be modelled using a-priori suppositions on the meridional flux variations or stochastic parameterizations (Choudhuri and Karak, 2009; Karak, 2010; Usoskin et al., 2009). The general dynamo theory cannot naturally reproduce the quasi-cyclicality occurrence of GM-like events; for that, some prescribed changes in the dynamo parameters are also required (Charbonneau and Dikpati, 2000). Therefore, determining the possible celestial influence in the modulation of solar magnetic activity is important for a wide spectrum of disciplines, for understanding how the solar dynamo works and for elucidating the characteristics of the mechanism behind the Sun-Earth relationship.

In this work we looked for a causal link in the planetary hypothesis of solar cycle. We continue the line of inquiry related to Wood and Wood (1965) and Fairbridge and Shirley (1987). Nevertheless, we search for particular changes in Sun acceleration and concentrate on long term variations of the solar cycle (the prolonged minima). The aim was to find and quantify some distinctive dynamics in planetary forcing related to solar barycentric movement and its possible relationship with these prolonged minima.

We show that there were unique configurations in Sun acceleration related to Sun-barycentre geometry around 1632, 1810 and 1990. They are produced by the well-known planetary quasi-alignments and are the cause of the retrograde SIM intervals. In particular, in these episodes, the radial component of the Sun's acceleration (i.e., in the barycentre-Sun direction) had an exceptional magnitude. These radial impulses in Sun's velocity occurred at the very beginning of MM, during DM and also at the maximum of cycle 22, before the present extended minimum.

The Sun is in a free fall state (see e.g., Shirley, 2006), therefore an underlying physics that explains the possible consequences of this barycentric motion in the internal structure of the Sun is needed.

Recently, Wolff and Patrone (2010) presented a new theory of Sun-planets interaction that is sensitive to Sun-dynamics with respect to the barycentric reference frame and could influence the solar cycle through the storing and releasing of potential energy in the solar interior. We apply our results in this novel theory and found a very important effect on the Sun´s capability of storing this hypothetical PE. This process (that lasts for ~80 yr.) begins about 40 years *before* the angular momentum inversions, around 1593, 1772 and 1951; i.e., before MM, DM, and before the present extended minimum. We also found a strong correlation between planetary torque and the sunspots international number one Schwabe cycle before and after the radial impulse seen around 1990. Our conclusions suggest a dynamical characterization of peculiar prolonged solar minima. We discuss the possible implications of these results for the solar cycle including the present extended minimum.

## 2. Three dimensional solar inertial motion and dynamical quantities

The SIM is reproduced by the integration of 3D equations of motion of the solar system planets (from Mercury to Neptune) which is performed by using the Mercury6.2 program (Chambers, 1999) in the high precision Bullirsch-Stoer mode. This is a state of the art ForTran code widely used in celestial mechanics and planetary system formation simulations. The planetary positions and velocities obtained were used to calculate the position **r**=**r**(*t*) , velocity **v**=**v**(*t*) and acceleration **a**=**a**(*t*) of the Sun in the ecliptical barycentric J2000.0 reference system, hereafter "the inertial system". Furthermore, we calculated a set of derived dynamical quantities such as the orbital angular momentum (**L**), its variations (d**L**/d*t*; d$^2$**L**/d*t*$^2$; etc.) and other interesting dynamical parameters. These quantities, were calculated using the "astronomical" units system: astronomical unit (AU) for distances; solar mass (Ms) for masses and days (d) for time. In this system of units the Sun's mass is obviously 1 and so we omit this value in what follows. The orbital angular momentum is:

$$\mathbf{L} = \mathbf{r} \times \mathbf{v} \tag{1}$$

and its components:

$$\begin{aligned} L_x &= (y\, v_z - z\, v_y) \\ L_y &= (z\, v_x - x\, v_z) \\ L_z &= (x\, v_y - y\, v_x). \end{aligned} \tag{2}$$

The planetary torque **τ** is defined, also as is customary, as the derivative of the **L** vector:

$$\begin{aligned} \tau_x &= (y\, a_z - z\, a_y) \\ \tau_y &= (z\, a_x - x\, a_z) \\ \tau_z &= (x\, a_y - y\, a_x), \end{aligned} \tag{3}$$

where $a_x$, $a_y$ and $a_z$ are the components of solar acceleration in the inertial system. The rate of change of $L$, $T = dL/dt$ (the so called *planetary torque* in most of the literature), is easily calculated as:

$$T = \frac{1}{L}\boldsymbol{\tau} \cdot \mathbf{L} = \frac{1}{L}\left((y\, a_z - z\, a_y)L_x + (z\, a_x - x\, a_z)L_y + (x\, a_y - y\, a_x)L_z\right). \tag{4}$$

We have also calculated all these dynamical quantities using the *invariable* barycentric system as the inertial system; it has the invariable plane of the Solar System as the fundamental plane (Burkhardt, 1982), looking for possible distortion effects in *x-y* projected quantities, due to the solar movement being close to this plane (Jose, 1965); but we have not found any noticeable effect in the calculated dynamical parameters.

### 2.1 *Retrograde Solar Motion*
Our analysis begins with the study of well-known angular momentum inversions around 1632 and 1811, because they were sporadic and important events in the Sun barycentric dynamic and are related to MM and DM, respectively (Jose, 1965; Fairbridge and Shirley, 1987; Javaraiah, 2005). The only known was that the *z*-component of **L** ($L_z$) becomes weakly negative at these moments, the Sun's movement around the barycentre was retrograde (clock-wise) and around these moments the giant planets were quasi-aligned in a kind of superior conjunction, i.e., Jupiter was placed, with respect to the other giants, on the opposite side of the Sun (with a longitude difference about 180 deg.) (Javaraiah, 2005).

Therefore, in order to examine the spatial evolution of *L*, we calculated the inclination (*i*) of the osculating solar orbit (OSO). The OSO is normal to **L** by definition. But we also calculated *i*, because Jose (1965) did not find appreciable variations of the OSO inclination which seems to be strange because the angular momentum inversions were indeed detected. Then we calculated, using celestial mechanics equations, the inclination of the OSO:

$$i = \cos^{-1}\left(\frac{L_z}{L}\right); \quad (5)$$

the OSO ascending node:

$$\Omega = \tan^{-1}\left(-\frac{L_y}{L_x}\right); \quad (6)$$

and the position angle *u* of the Sun in the OSO measured from the nodal line (often called *argument of latitude*) that in the general case, *i* and $\Omega \neq 0$, is:

$$u = \tan^{-1}\left(\frac{z}{\sin i(x \cos\Omega + y \sin\Omega)}\right). \quad (7)$$

## 3. Results

The results of SIM simulations show the well-known angular momentum inversions to be consistent with the retrograde solar motion, around 1632 (1632.31 to 1632.92), 1811 (1810.79 to 1812.02) and in 1990 (1989.68 to 1990.91). The evolution of OSO inclination around a particular retrogression time (1811) is shown in Fig. (1a). It is clearly seen that *i* passes 90º at the moment of retrogression and consistently, **L** vector librates at the same angle with respect to its initial position, near coincident with z-direction (Fig. 1b). These librations take about 1 year. Therefore, this new result shows that the angular momentum inversion is a gradual but rapid process. As was mentioned, giant planets quasi- alignments occur, and they are considered responsible for these retrogressions (Javaraiah, 2005). However in our simulations, in the last millennium, similar planetary configurations have occurred around 953 – 991; 1132 – 1170; 1310-1349; 1451 – 1498; i. e., each ~179 yr. On the other hand, the alignments of 1632, 1811 and 1990 were quite different (very aligned), therefore raising the question about the forcing mechanism responsible for these changes in *L*, or about any particular or impulsive force acting in the Sun at these inversion epochs. It is not a trivial question. For example, in a keplerian orbit, a change in angular momentum direction is only possible if a perturbing acceleration is acting *normal* to the orbit.

The modulus of the Sun´s acceleration (not shown for brevity) has an oscillating-sinusoidal behaviour, we found no distinctive features related to these epochs. Neither did we find any noticeable effect in the Cartesian components of solar acceleration in the inertial system. These sinusoidal behaviours are due to the quasi-periodic movement of the planets, and are dominated by Jupiter's orbital period.

Nevertheless, there is the possibility that in other directions, for instance one more related to the Sun barycentric dynamic, a component of solar acceleration might exhibit a distinctive manifestation, and this is not a trivial issue either. For example, it is very interesting to address how the *radial* component of solar acceleration (i.e., in the barycentre-Sun direction) evolves, especially taking into account the planetary quasi-alignments. The radial direction evolves with the Sun's orbital motion; therefore, we were motivated to look for the component of the acceleration projected in an *orbital* (i. e., attached to OSO plane) system, this system rotates with the Sun`s mean motion and shares all the "irregularities" of OSO. The use of this system is very convenient because it permits us to obtain the radial component of solar acceleration, and, at the same time, we can obtain the projection of the orbital angular momentum with respect to the acceleration, which is an important dynamical reference. Moreover, this transformation was very straightforward because we already had the necessary quantities involved in the transformation between both reference systems (inertial and orbital), as is explained as follows. The orbital system is defined by the radiovector direction (**r**), which always points from the barycentre toward the Sun; the normal to the orbit (**h**), it is always coincident with **L** direction; and the transversal direction (**t**) that completes the orthogonal directions of the new system (Fig. 2), the origin of this new system is the same as the inertial system. Therefore, the Sun´s acceleration components in this non-inertial system are ($a_r$, $a_t$, $a_h$); and are obtained by means of the following matrix multiplication:

$$(a_r, a_t, a_h)^t = \mathbf{Rz}(u)\ \mathbf{Rx}(i)\ \mathbf{Rz}(\Omega)\ (a_x, a_y, a_z)^t, \quad (8)$$

where **Rz**, **Rx**, are the rotation matrices around the indicated axes; these matrices have as arguments all the angular elements already calculated in Sec. 2.1 (Eqs. 5,6 and 7).

The idea of this transformation was simple: to obtain the projections of inertial acceleration in these particular directions, looking for sudden changes or conspicuous behaviour (but in a simple way, instead of performing more calculations to obtain the direction cosines between each involved quantity). Obviously, if you write the equation of motions in this non-inertial frame, the fictional forces appear in the description of the acceleration (i.e., Coriolis, centrifugal, and also, the Euler term, because the system has angular acceleration).

We found an "impulsive" manifestation of $a_r$[1] and $a_h$ at epochs of orbital inversion, but not in $a_t$ component, which shows an oscillatory behaviour. First of all, for a complete description of the transformed acceleration that takes into account all components but preserves this impulsive manifestation, we sum $a_t$ and $a_r$ components to obtain the acceleration in the Sun's orbital plane, i. e., the strength of the *coplanar* component of solar acceleration, $a_\pi = (a_t^2 + a_r^2)^{1/2}$. This component and also $a_h$, that we call the *normal* component of solar acceleration, is shown in Fig. 3 from 800-2100 A.D. As was explained, this impulsive behaviour of $a_\pi$ is due to $a_r$ which during the retrograde motion intervals, displays sudden changes, very much larger than the average of the near values; moreover, $a_r$ component that is always negative, becomes positive. This is due to the barycentre being left "outside" the solar orbit, i.e., the Sun fails to loop the barycentre (see e.g., Jose, 1965, Fig. 1d in that paper).

The normal component ($a_h$) also has an exceptional increase at these epochs, which is explained as follows: When the OSO becomes retrograde, the angular momentum is inverted and this inversion aligns the **L** vector towards the planetary acceleration direction. Therefore, *$a_h$ is not an impulsive change in planetary acceleration normal to the solar orbit*. This must be seen as the maximum projection of **L** in the solar acceleration direction, but in a contrary sense. Obviously, changes in this component are due to orbital libration, not to an important increase in the *z*-component of acceleration in the inertial system, because planetary acceleration is always near the ecliptical plane. But these features in $a_h$ are not trivial, they mean that at the times of angular momentum inversions, **L** is almost anti-parallel to the **a** vector for a while. Indeed, the angle between **L** and **a** oscillates between ~130-179 deg at these epochs.

We highlight that these radial impulses never occurred before 1632 during the last millennium; whereas similar planetary configurations did happen, as was already mentioned. This means that these impulses only arise at epochs of orbital inversions and they are truly responsible for these particular phenomena. Motivated by these findings, we performed a phenomenological comparison with the corresponding MM and DM events. To that end, we used the sunspots physics-based reconstruction by Solanki et al. (2004). In addition, we added the monthly averaged international sunspot group number observed from 1795 until present, taken from NOAA Solar Data Center
[ftp://ftp_ngdc.noaa.gov/STP/SOLAR_DATA/SUNSPOTS_NUMBERS](ftp://ftp_ngdc.noaa.gov/STP/SOLAR_DATA/SUNSPOTS_NUMBERS)).
We also used this data smoothed with a low pass Savitsky-Golay filter (Press et al., 1992)
to remove "noisy signal" or minor spikes.

Fig. 4 shows the coplanar and normal components of the acceleration showing the particular dynamics around 1632, 1811 and also around 1990. Therefore, the periods of $L_z$ inversion were characterized by a unique geometry of planetary acceleration with respect to the Sun-barycentre system, i.e.: the radial component of solar acceleration had an exceptional magnitude and the **L** vector was anti-parallel to acceleration. This has occurred in the last millennium only at the epochs of MM, DM and at the maximum of cycle 22, around 1990.

**3.1** *More phenomenology*
It is very suggestive that the first detected impulse occurs at the very beginning of the MM, the second one occurs in the middle of the DM, and the last impulsive event was coincident with the maximum of the cycle 22.

Taking into account these facts, it is straightforward to consider the possibility of finding some phenomenological correlations between the barycentric dynamic of the Sun and solar cycle around 1990, as have been previously reported for other times (e.g., Jose, 1965; Wood and Wood, 1965; Landscheidt, 1999).
We plot in Fig. 5, the OSO inclination, *T*, $L_z$ and the monthly observed and filtered sunspots numbered series (SN). We can see an apparent phase synchronization (correlated, anti-correlated) between SN and *T*. The correlation between *T* (only giant planets) and the observed SN taking into account a Schwabe cycler before the radial impulse is 0.76, and after the impulse is –0.52. Notably, the duration of the maximum of the smoothed cycle is coincident with the retrograde motion interval.

The current exceptionally long minimum of solar activity has attracted attention with respect to the possibility of developing a new grand minimum in the next decades (Badalyan et al., 2001; Livingston and

---
[1] See Fig. 6 in the last section for a graph of $a_r$.

Penn, 2009; Russell et al., 2009; Tan, 2010; Ahluwalia and Ygbuhay, 2012). Moreover, suggestions of a new imminent GM were taken by Feulner and Rahmstorf (2010) and Song et al. (2010) to evaluate the possible effect on present Earth climate. Furthermore, de Jager and Duhau (2009) used non-linear analysis of past solar dynamo inferred activity, to conclude the existence of a transition point between two different states after 1990 that could produce a new minimum after this epoch. It is interesting to note that in 1990 there was another G-O rule violation involving cycle 22 and 23 (Javaraiah, 2005; Kane, 2008), i. e., before the radial impulse at the maximum of the cycle 22. Nielsen and Kjeldsen (2011) analysed the spotless days of solar cycle and concluded that the ongoing accumulation of spotless days is comparable to that of cycle 6 near the DM, among other cycles.

The current prolonged minimum is in progress, therefore, its very nature has not been confirmed yet. The other two prolonged minima related to these impulsive accelerations were GM events. We will see in the next section that possible dynamical implications of these impulses for the solar cycle begin before the onset of these GM events and before the onset of the present extended minimum. Therefore; the aforementioned observational facts and our findings should support and encourage attempts to study the present solar minimum in relation with a possible forthcoming GM event.

## 4. Discussion and hypotheses

Physical mechanisms of Sun-planets interaction, that involve the solar interior, are not obvious; moreover, quite elemental errors trying to link solar particles and SIM have been made (Shirley, 2006). Nevertheless, a spin-orbit coupling hypothesis was proposed to explain the phenomenology between $L$ variations and solar activity through solar rotation and dynamo functioning (Blizard, 1981; Jucket, 2003; Javaraiah, 2005; Shirley, 2006; Perryman and Schulze-Hartung, 2011). The idea is that planets can transfer orbital momentum to the Sun's rotational angular momentum, and this variation could interact with the solar dynamo through a putative mechanism. Indeed, the solar angular momentum exhibits great variations, whereas the planet's angular momentum show variations as much as 0.5 % in about 20 yr, $L$ has variations up to 20%.

Therefore, some part of $L$ could be transferred to spin angular momentum; but the inverse can also be true as was shown by Javaraiah (2005, Sec.4).

We have found at the times of angular momentum inversions, **L** is almost anti-parallel to the **a** vector for a while. Indeed, the angle between **L** and **a** oscillates between ~130-179 deg at these moments. Therefore these configurations must be investigated in the frame of the planetary hypothesis taking into account the spin-orbit coupling as working mechanism or line of inquiry (Javaraiah, 2005).

The Sun is not a "perfect sphere" (e.g., Rozelot et al., 2009), and is torqued by the planets and precesses in the inertial space. Authors such as Javaraiah (2005) and Juckett (2000, 2003) have provided evidence of spin-orbit coupling in the Sun; therefore it is a reasonable hypothesis that must be corroborated. But, whereas this precessional effect is already observed (Fränz and Harper, 2002), there is no model explaining the suspected angular momentum transference. Also Paluš et al. (2007) have shown by a qualitative analysis of the radius of curvature of the solar barycentric orbit that this parameter and sunspots series have been synchronized in the past. Nevertheless, this does not imply that both quantities share the same physical mechanism; but this could be an effect of collective synchronization of oscillatory systems (Strogatz, 2001). Therefore, the particularities of SIM and solar rotation are an intriguing question not easy to address.

We have found impulses in the Sun's acceleration in the radial direction that are the cause of orbital inversion that place the **L** vector anti-parallel to the planetary acceleration direction.
We have also shown an interesting phenomenology between these forcing configurations and solar activity related to special events in long term solar activity. These forcing configurations cannot say anything by themselves with respect to possible effects in the solar interior; for that, a theory linking planetary movements and the modification of the physical parameters of the solar interior is necessary. Recently Wolf and Patrone (2010) presented the first working mechanism devoted to explain modifications of stars interiors by planetary gravity. They showed that a "cell" inside a rotating star with orbiting planets (i. e. with measurable *inertial* motion), can create potential energy per unit mass (PE) that could be released by internal processes (i.e., by natural convection).

The mechanism proposed by Wolff and Patrone (2010) (hereafter WP) depends on the barycentric velocity and position of the Sun. These authors have also shown that the strongest case is the "vertical" one, i.e., the larger storage occurs when the cell is near the Sun´s centre-barycentre direction (Fig. 3 in that paper). It is worth noting that WP had not found any mechanism directly linking solar cycle activity and planetary movement; therefore, a full theory that would physically explain how the SIM may influence the solar cycle has not been developed yet (as was already stated in Sec. 1). Nevertheless, the authors suggest that this mechanism might explain some of the many reports of correlations between the strength of solar activity and planetary motions and propose evidence of this. For example, in the named vertical case, energy could be transferred into solar (gravity) g-mode oscillations which would provide rapid upward energy information transport. They argue that a full or partial overturning of a cell in the Sun would generate a field of turbulence near the cell boundary and throughout the cell. As the turbulence decays, heat would be deposited in and near the cell. This suggests, as a test, an increased solar activity when cell overturning or its generated heat stimulate extra near-surface convection. The mechanism, in general, could induce a strengthening of solar activity (the more energy available, the greater the solar activity).

Full calculations taking into account all the cases and its variations in WP theory is clearly beyond the scope of this paper; nevertheless, to show if our detected impulses in acceleration (clearly visible in the radial component) have any influence in this proposed PE storage, we calculate in the light of this new theory, and using the barycentric positions and velocities found in our simulations, the specific potential energy stored in the vertical case (*PEv*).

Fig. 6, shows the radial component of the acceleration $a_r$, *PEv* stored at 0.16 solar radius facing the barycentre and the observed sunspots series SN. As **v** and **r** vary, *PEv* departs from the calculated values in WP (Fig. 3 in that paper) *PEv* values oscillate with a period of approximately 20 yr, due to Jupiter-Saturn conjunctions, but remain approximately bounded between 1-10 $m^2 s^{-2}$, except when the Sun approaches retrograde motion intervals (i.e., impulsive changes in Sun velocity).

Around 1590 *PEv* drops approximately one order of magnitude with respect to its average neighbours values; then in 1632 the fall is about two orders of magnitude; later, around 1670, *PEv* has another big drop. Then the *PEv* signal recovers its oscillatory-bounded values. This configuration is reproduced (but with different intensity) around the other two epochs of radial impulses (these minima have been marked with an 'X' symbol in the Fig. 6). Therefore we found that, in the studied period, the most important perturbations to PE storage in the WP model, begin *before* these last GM events and *before* the impulse of 1990. The relationship of this behaviour with radial impulses is explained as follows.

The Saturn-Uranus synodic period is about 40 years; therefore, 40 years before (i.e., ~1593, 1772, 1951) and 40 years after (i.e., ~1670, 1850, 2030) the giant planets quasi-conjunction, an increase of $a_r$ arises, because these planetary configurations are quasi-repeated around those other times (Fig. 6). The *PEv* minima are more or less pronounced depending on the relative positions and velocities of the planets in those other times (the planetary periods are non-commensurable). Therefore, in the WP model, the dynamical effect of the planetary alignments is not as impulsive as one would think, but rather takes about 80 years due to the approximate repetitions of the planetary quasi-conjunctions.

We ignore the precise effect of these long-term variations of PE inside the Sun in the context of WP theory, but if this theory is basically correct, certainly should be one effect. Our findings should stimulate observational search for these possible effects and the development of accurate models to explain their influence on the solar cycle.

We have shown that epochs of radial impulses are related to prolonged minimums in solar activity.

Based on our dynamical findings and taking into account these theoretical results and the observational evidence that suggest that this solar minimum is a very conspicuous one, we can make the hypothesis that MM, DM and the extended minima after cycle 22, shared certain changes due to planetary accelerations that might have a significant effect in the solar interior through variations to the proposed PE storage.

Therefore, we propose a dynamical characterization of peculiar prolonged minimums: two of them (MM and DM) are *bona-fide* gran minima events, but the last is in progress and its nature has not been confirmed yet.

We also can speculate the existence of a putative mechanism linking these hypothetical variations in PE and solar dynamo governing parameters; i. e, if these big drops in PE related to radial impulses really exist, they might contribute to a triggering mechanism of a slow-down in the solar cycle.

Therefore, we could propose that the occurrence of obvious radial impulses on Sun velocity have associated these types of prolonged minima events; but certainly, this cannot be seen as a necessary condition.

If these perturbations due to impulsive accelerations occur, they certainly occur with others of different nature, both internal (e.g., magnetic instabilities, variation of meridional circulation speed, inherent to dynamo dynamics, etc.) and external (e.g., tidal effects, spin-orbit coupling ?, etc.) to the Sun. Since the solar dynamo has very non-linear dynamics (Duhau and de Jager, 2008), the resulting effect is very difficult to predict.

In other prolonged minima, such as Spörer minimum (1415-1534), when these strong impulses were absent, these different kinds of perturbations might have had greater importance, or the dynamo may have been in an appropriate initial state likely to produce a GM event.

Regarding the beginning of this new prolonged minimum, it is important to note that other physical facts support the hypothesis of a new possible GM after cycle 22; for instance, variations in the Sun's meridional flow. Indeed the strength of the polar fields produced after the cycle maximum in 2000–2001 was only about half that seen in the previous three solar cycles (Hathaway and Rightmire, 2010) and this is a characteristic that is shared with MM (Karak, 2010). Furthermore, cycle 24 started much later than average. The late start for cycle 24 followed a long quiet minimum unlike any in the past 100 years. Also space-based measurements indicate that the solar EUV irradiance was also anomalously low during this time or at least lower than the previous solar minimum (Badalyan, et. al, 2001). Ionospheric observations find that the ionosphere was lower and cooler than typical solar minimum conditions, and results from analyses of satellite orbital elements show that the upper thermosphere was significantly less dense, i.e., cooler, than the previous several solar minima, and indeed cooler than any other such period on record since the beginning of space flight (Solomon et al., 2010, and references therein). For this work, the exact nature of the current prolonged minimum (i.e., if it is a GM event) is of secondary importance: the central result is that the three last prolonged minima can be related by common barycentric dynamics and, as a hypothesis, this dynamics might be a perceptible effect in Sun's interior, plausibly related to the solar cycle.

Undoubtedly, a precise classification of GM events requires the study of more events, for which the next cycles will be a strong test of all these hypotheses and fundamental for the classification of this prolonged minimum. As was mentioned in Sec. 1, only a few theoretical models with *a priori driving forces* can "naturally" reproduce GM-like events. Consequently, it is important to evaluate in the frame of WP theory, the effect of impulsive changes in acceleration with relation to dynamo functioning (e.g., turbulence variations), and this is work in progress.

The planetary hypothesis of solar cycle has received little attention from the astrophysical community, we think, because it is accustomed to addressing these kinds of problems from a "classical pertubative" approach. The planetary accelerations, as we have shown, are small quantities, consequently, it is difficult to imagine that the planets might affect the solar interior. Nevertheless, the theory of WP overcomes this difficulty, and taking into account the position of several authors (e. g., Grandpierre, 1996; Shirley, 2006; Paluš et al., 2007; Scafetta, 2010; Tan, 2010) and our own opinion, the chaotic nature of the solar dynamo can amplify the effects of a weak external periodic forcing through resonances, collective synchronization and feedback mechanisms. In addition, the Sun is in a plasma state and the plasma system is always very fragile, even with a very small perturbation, a variety of instabilities very easily develop and they can be a triggering mechanism of variations and cycles.

## 5. Conclusions

In this contribution, we have revisited the planetary hypothesis of solar cycles, i.e., the idea by which the planets of the solar system might have some influence on solar magnetic activity.

We searched for particular changes in Sun acceleration and concentrate on long term variations of the solar cycle looking for a possible causal link or physical connection between Sun barycentric dynamics and solar activity related to prolonged solar minima.

Using an orbital reference system, we have found that, at the epochs of retrograde solar motion, the solar acceleration shows radial impulses (in Sun-barycentre direction), and the angular momentum inversions placed **L** vector in a quasi-anti-parallel direction to solar acceleration. Although giant planets quasi-alignments repeat every ~179 yr, it is only before MM, in the DM, and at the maximum of cycle 22, before this extended solar minimum, that this particular barycentric dynamic is shared. We have clarified the nature of planetary forces acting at these epochs and have showed their possible effects in a new theory of Sun-planets interaction.

First of all, these findings are very important with respect to the planetary hypothesis, because never before have distinctive accelerations in the Sun's barycentric dynamics been reported; and neither have these kinds of forcing been related to particular events in the solar cycle.
Secondly, our findings are important with respect to a new theory of Sun-planets interaction, since this theory links *PE* variations with solar internal (possibly with dynamo) functioning.
Whereas these impulses are "the evident linkage" in this matter, their associated dynamics is more complex, and the possible dynamical effects (following WP theory) begin 40 yrs. before the evident radial impulses, .i.e., before MM, DM, and before the present extended minimum.
Thus, we argue for a classification or dynamical characterization of certain prolonged solar minima and also in favour of encouraging attempts to study the present prolonged minimum in relation with a possible forthcoming GM event. Our findings should stimulate observational search for these possible effects in the frame of WP theory and the development of accurate models to explain their possible influence on the solar cycle.

## Acknowledgments

The authors acknowledge the support of grant PID-UTN 1351 "Forzantes externos al planeta y variabilidad climática" of UTN; and grant PICT – 2007 – 00438 of ANPCyT of Argentina.

FIGURE CAPTIONS

Fig. 1 – Panel a): evolution of OSO inclination in a particular retrogression epoch (1811). It is clearly seen that *i* passes 90º at the moment of retrogression. Panel b): *i* vs. $L_z$ plane showing the **L** vector libration at the times of orbital retrogression. Around 1632 the libration is less pronounced.

Fig. 2 – The orbital system is defined by the radiovector direction (**r**), which always points from the barycentre toward the Sun; the normal to the orbit (**h**), it is always coincident with **L** direction; *i* is the OSO inclination; *u* is the position angle of the Sun in its instantaneous orbit; Ω is the longitude of the ascending node of OSO orbit.

Fig. 3 – Coplanar and normal components of solar acceleration in the interval 800-2100 A.D: first panel, normal component $a_h$; second panel, coplanar $a_\pi$ component. The OSO inclination *i* is also shown. Accelerations reckoning in UA d$^{-2}$.

Fig. 4 – Coplanar and normal components of solar acceleration. Their impulsive nature after the MM (M), around 1632; in the middle of DM (D), in 1811; and at the maximum of cycle 22, around 1990, is explained in the text. SN are the sunspots series observed-filtered (dashed line) and the reconstruction of Solanki et al. (2004) (solid line). Accelerations reckoning in UA d$^{-2}$.

Fig. 5 – OSO inclination *i*; $L_z$, d$L$/d$t$ (only giant planets); and the monthly observed (solid line) and observed-filtered (dashed line) sunspots series SN around 1990. Physical units: Ms, UA and d.

Fig. 6 – Radial acceleration $a_r$; PE storage at 0.16 solar radius facing the barycentre (*PEv*), and the observed sunspots series (SN). The particular dynamics around the impulsive events produce a dramatic decrease of the hypothetic PE in the studied solar zone. The 'X' symbol marks the minima in *PEv* values; they occur ~1593-1632-1670; ~1772-1811-1850; ~1951-1990-2030 (see the text for explanation). Acceleration reckoning in UA d$^{-2}$. *PEv* plot in logarithmic scale.

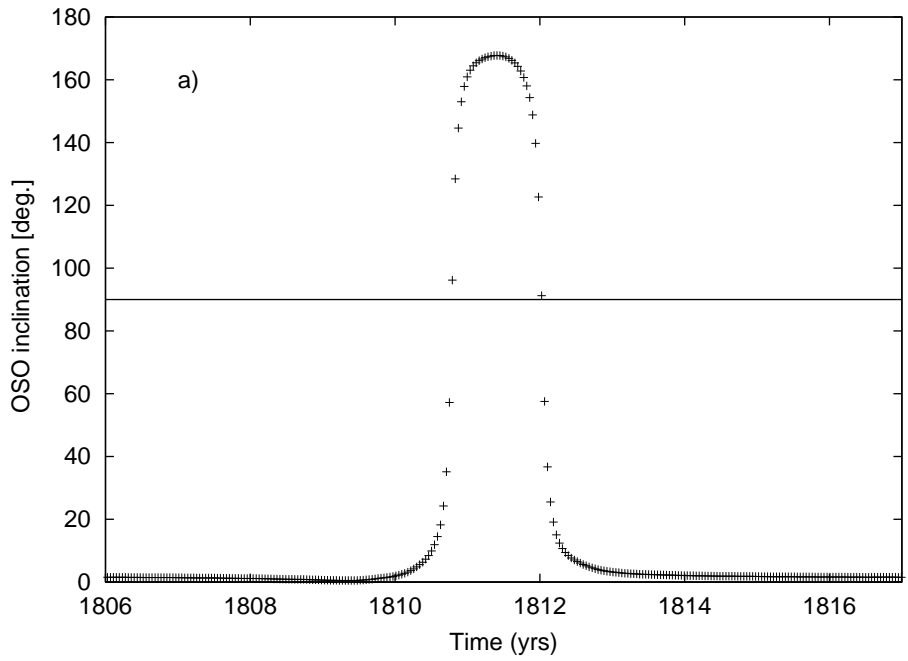

Fig. 1 a

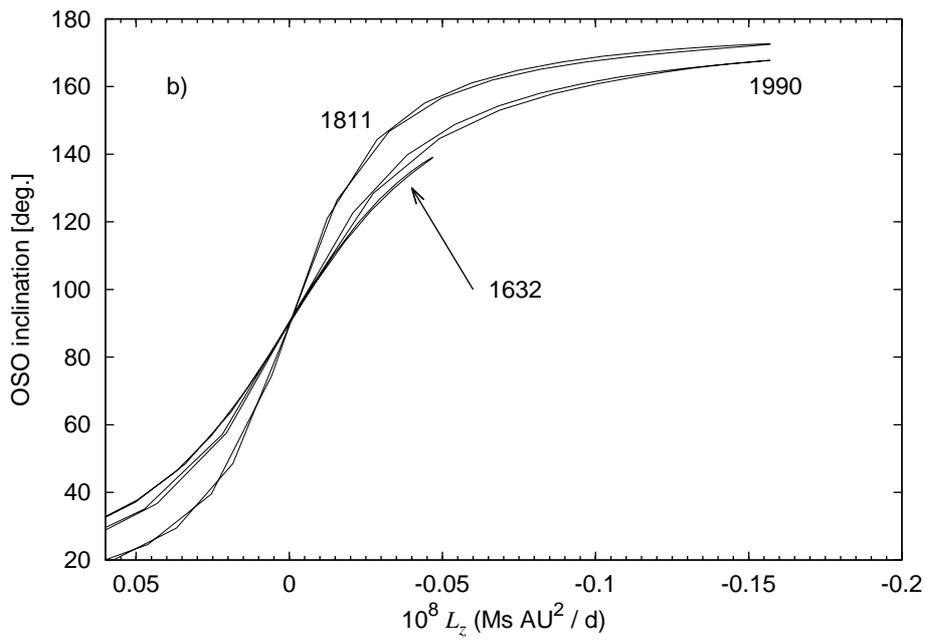

Fig.1b

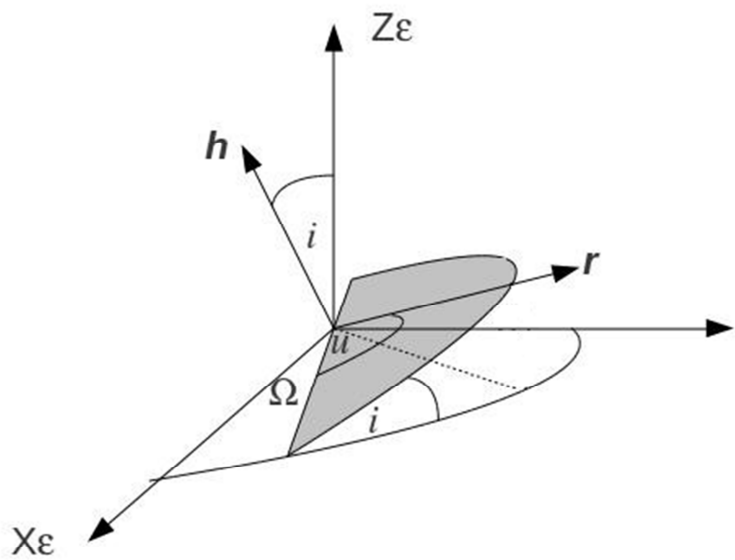

Fig. 2

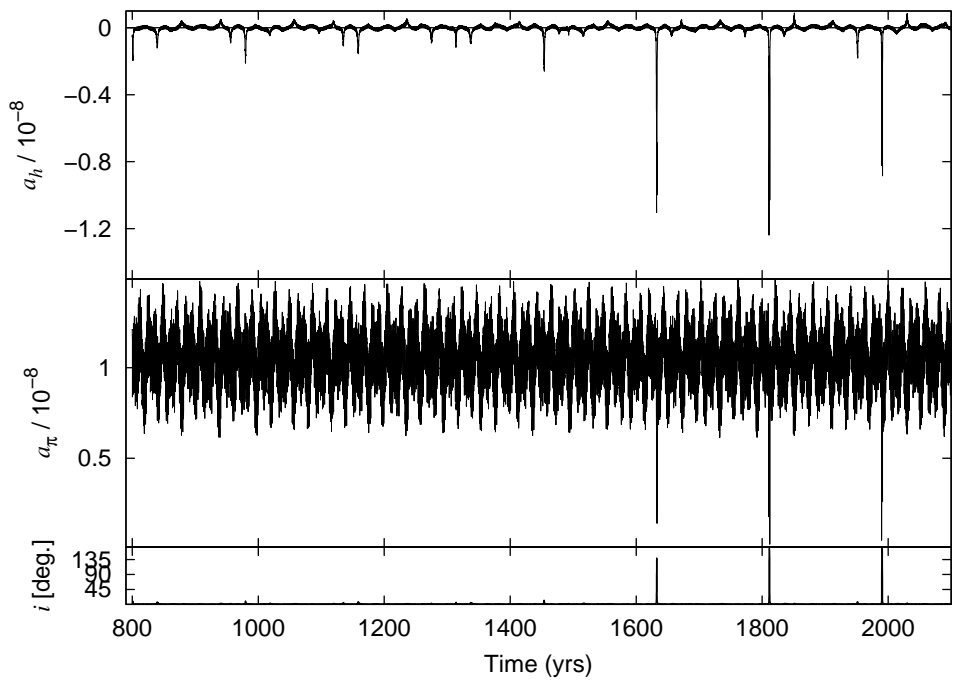

Fig. 3

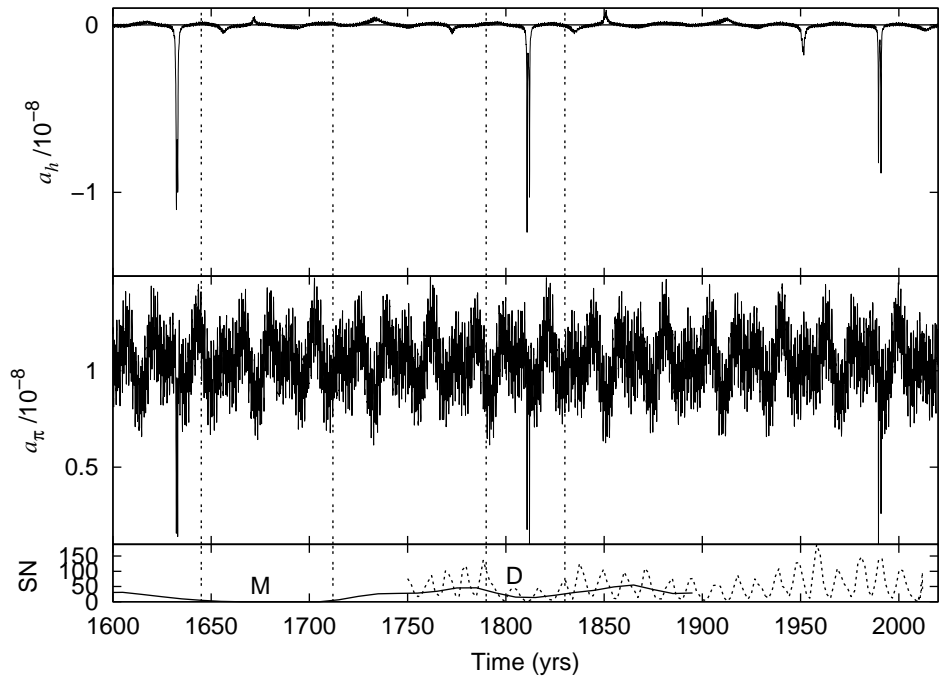

Fig. 4

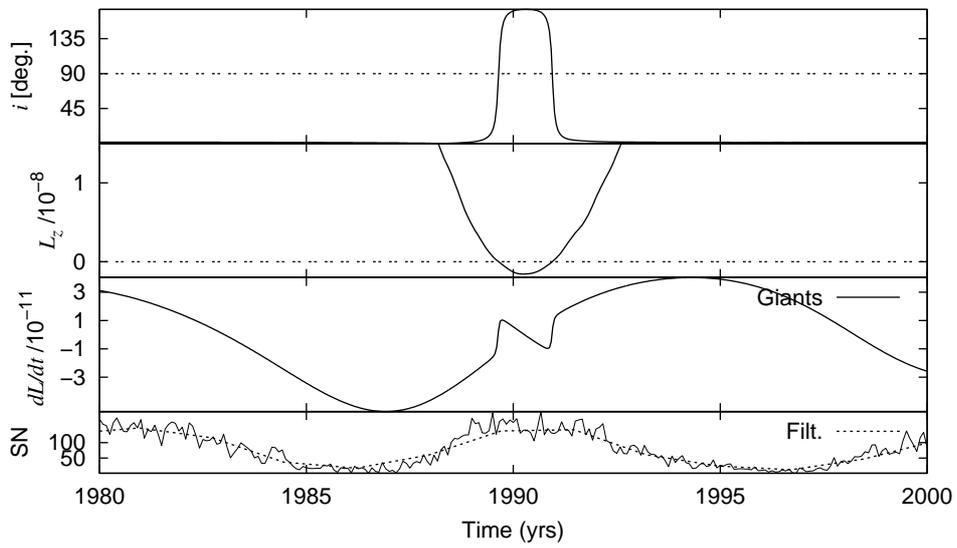

Fig.5

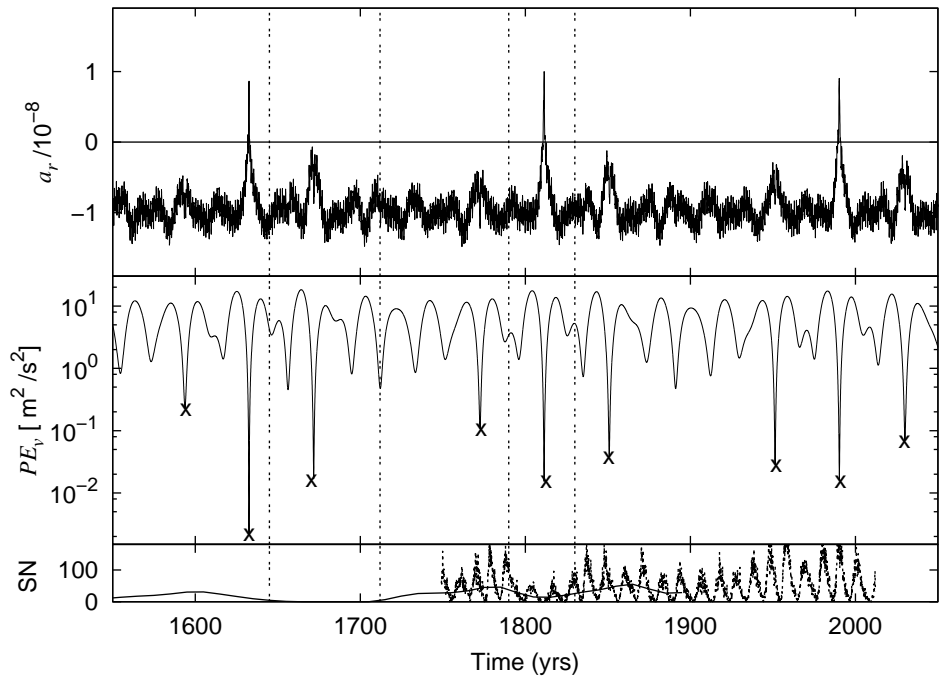

Fig. 6